\documentclass[aps,prc,twocolumn,superscriptaddress,showpacs,floatfix,nofootinbib]{revtex4-1}
\usepackage{amsmath,graphicx,color}
\begin{document}

\title{
    $p_T$-Dependent Particle Number Fluctuations From Principal Component
    Analyses in Hydrodynamic Simulations of Heavy-Ion Collisions
}

\author{Fernando G. Gardim}
\affiliation{
    Instituto de Ci\^encia e Tecnologia, Universidade Federal de Alfenas,
    37715-400 Po\c cos de Caldas, MG, Brazil
}
\affiliation{
    Institut de physique th\'eorique, Universit\'e Paris Saclay, CNRS, CEA,
    F-91191 Gif-sur-Yvette, France
}
\author{Fr\'ed\'erique Grassi}
\affiliation{
    Instituto de F\'{i}sica, Universidade de S\~{a}o Paulo, 
    R. do Mat\~{a}o 1371, 05508-090 S\~{a}o Paulo, SP, Brazil
}
\author{Pedro Ishida}
\affiliation{
    Instituto de F\'{i}sica, Universidade de S\~{a}o Paulo,
    R. do Mat\~{a}o 1371, 05508-090 S\~{a}o Paulo, SP, Brazil
}
\author{Matthew Luzum}
\affiliation{
    Instituto de F\'{i}sica, Universidade de S\~{a}o Paulo, 
    R. do Mat\~{a}o 1371, 05508-090 S\~{a}o Paulo, SP, Brazil
}
\author{Jean-Yves Ollitrault}
\affiliation{
    Institut de physique th\'eorique, Universit\'e Paris Saclay, CNRS, CEA,
    F-91191 Gif-sur-Yvette, France
}

\begin{abstract}
We carry out a principal component analysis of fluctuations in a hydrodynamic
simulation of heavy-ion collisions, and compare with experimental data from the
CMS collaboration. The principal components of anisotropic flow reproduce the
trends seen in data, but multiplicity fluctuations show a difference in
transverse momentum dependence. We construct an analytical toy model and verify
that hydrodynamic simulations agree with its predictions. The difference in the
momentum trend is likely due to larger fluctuations in transverse momentum of
hydrodynamic models than seen experimentally.
\end{abstract}

\maketitle

\section{Introduction}

The expansion of the matter formed in nucleus-nucleus collisions at relativistic
energies produces a collective transverse flow. This flow is the response to
the density gradients in the initial fireball. It is azimuthally asymmetric
because the initial fireball is anisotropic and contains hot spots. These
inhomogeneities are of interest: they reflect the poorly known mechanism of
energy deposition, via the strong interaction, when two nuclei collide, and
their influence on the final flow depends on fluid properties, which are also
poorly known (e.g. shear and bulk viscosities). A lot of work has been done to
relate initial inhomogeneities and final flow of produced particles. In
particular the mapping between initial conditions and anisotropic flow has been
studied globally and event-by-event \cite{Gardim:2011xv,Gardim:2014tya,
Niemi:2012aj,Niemi:2015qia,Fu:2015wba,Noronha-Hostler:2015dbi}. To get more
detailed information on fluctuations in the initial state, a useful observable
is the factorization breaking ratio~\cite{Gardim:2012im,Kozlov:2014fqa,
Heinz:2013bua,Khachatryan:2015oea,Gardim:2017ruc,Zhao:2017yhj,McDonald:2016vlt},
which encodes the correlations of flow harmonics at different transverse momenta
or pseudorapidities. More recently a new more precise tool was proposed, the
Principal Component Analysis (PCA) for event-by-event fluctuations
\cite{Bhalerao:2014mua,Mazeliauskas:2015vea,Mazeliauskas:2015efa} and first
experimental results for such an analysis have been presented by the CMS
collaboration \cite{Sirunyan:2017gyb}. The aim of this paper is to present a
hydrodynamical study of these observables and point out an interesting
difference between data and some hydrodynamic simulations for the $n=0$ leading
and subleading components, corresponding to multiplicity fluctuations. These
components are sensitive to physics not explored by anisotropic flow and can put
new constraints on initial conditions models, in particular on the transverse size of the fireball and its fluctuations.

\section{Principal Component Analysis}

Principal Component Analysis is a common technique for finding patterns in data
of high dimension. One tries to find new variables that incorporate as much as
possible of the variations. This amounts to diagonalizing the covariance matrix
(e.g. \cite{pcaref1}). It was first suggested to use it to study event-by-event
fluctuations in relativistic nuclear collisions in \cite{Bhalerao:2014mua}.
Consider a set of collisions or events. For each event, the single particle
distribution can be expanded as
\begin{align}
    \frac{dN}{d\vec{p}} & =
    \frac 1 {2\pi} \sum_{n=-\infty}^{+\infty} N(p_T) V_n(p_T)e^{-in\phi}
    \\ & =
    \sum_{n=-\infty}^{+\infty} \mathcal V_n(p_T)e^{-in\phi}
\end{align}
where $d\vec{p} = dy dp_T d\phi$, $\phi$ is the azimuthal angle of the particle
momentum. $\mathcal V_n(p_T)$ is a Fourier coefficient (without the usual
normalization by multiplicity) which is complex for $n\not=0$. Its magnitude and
orientation vary for each event. 

For each transverse momentum bin, the variance can be computed
$\langle |\mathcal V_n(p_T^a)|^2\rangle-|\langle \mathcal V_n(p_T^a)\rangle|^2$
(the average is performed over events) but brings no information about possible
relationship between different bins. To investigate how different bins are
correlated, one constructs the covariance matrix:
\begin{equation}
\label{defvndelta}
    \mathcal V_{n\Delta}(p_T^a,p_T^b) \equiv
    \langle \mathcal V_n(p_T^a)\mathcal V_n^*(p_T^b) \rangle -
    \langle \mathcal V_n(p_T^a) \rangle \langle \mathcal V_n^*(p_T^b) \rangle.
\end{equation}
The terms $ \langle \mathcal V_n(p_T^a)\rangle$ are zero by azimuthal symmetry,
except for $n=0$.

This covariance matrix is real, symmetric, positive-semidefinite. It can be
diagonalized and re-written in term of its real orthogonal eigenvectors
$\mathcal V_n^{(\alpha)}(p_T)$ 
\begin{equation}
    \mathcal V_{n\Delta}(p_T^a,p_T^b) =
    \sum_{\alpha} \mathcal V_n^{(\alpha)}(p_T^a) \mathcal V_n^{(\alpha)}(p_T^b),
    \label{eq:pca}
\end{equation}
from where one can express the flow vector in a given event as
\begin{equation}
    \mathcal V_n(p_T) =
    \sum_{\alpha} \xi_n^{(\alpha)} \mathcal V_n^{(\alpha)}(p_T),
    \label{eq:pcaV}
\end{equation}
where $\xi_n^{(\alpha)}$ are coefficients that vary from event to event
(specifically, uncorrelated, random complex numbers with zero mean and unit
variance). Terms in the right-hand side of Eq.~(\ref{eq:pca}) are ordered
according to the magnitude of the eigenvalues. Even by truncating the sum to the
first two or three terms, one typically obtains a very good approximation to the
left-hand side. The largest component $\mathcal V_n^{(1)}(p_T)$ is called the
leading mode, $\mathcal V_n^{(2)}(p_T)$ the subleading mode, etc. For
comparison with standard flow, it is useful to introduce the following scaled
principal components 
\begin{equation}
    \label{defscaled}
    v_n^{(\alpha)}(p_T) =
    \frac{\mathcal V_n^{(\alpha)}(p_T)}{\langle \mathcal V_0(p_T) \rangle}.
\end{equation} 
Once the dominant terms in Eq.~(\ref{eq:pca}) are determined (i.e. patterns are
found in our high dimension data), the physical meaning of these terms must be
investigated. This was done in~\cite{Bhalerao:2014mua,Mazeliauskas:2015vea,
Mazeliauskas:2015efa,Bozek:2017thv,Liu:2019jxg} and is discussed in section
\ref{sec:aniso} ($n=2,3$) and \ref{sec:mult} ($n=0$). 

\section{Results for anisotropic flow}\label{sec:aniso}

In this section and the next, we present results obtained from a hydrodynamic
simulation for a perfect fluid expanding in 3+1 dimensions starting from NeXus
initial conditions \cite{Drescher:2000ha}. The code used, NeXSPheRIO, has been
shown to lead to a consistent description of many flow data at top RHIC energies
\cite{Qian:2007ff,Andrade:2006yh,Andrade:2008xh,Andrade:2008fa,Gardim:2011qn,
Gardim:2012yp,Takahashi:2009na,Qian:2012qn}.

We also have some data accumulated for two centrality windows (0-5 and 20-30\%, where centrality is defined according to the number of participant nucleons) 
at $\sqrt{s}=2.76$ TeV and their compatibility with flow observables more subtly
related to fluctuations (scaled harmonic flow distributions, factorization
breaking ratio) has been tested \cite{RHICdist}. This code is therefore an
interesting tool for a first investigation of the PCA results obtained recently
by CMS at the LHC \cite{Sirunyan:2017gyb}. 

\begin{figure*}
\includegraphics[width=0.8\textwidth]{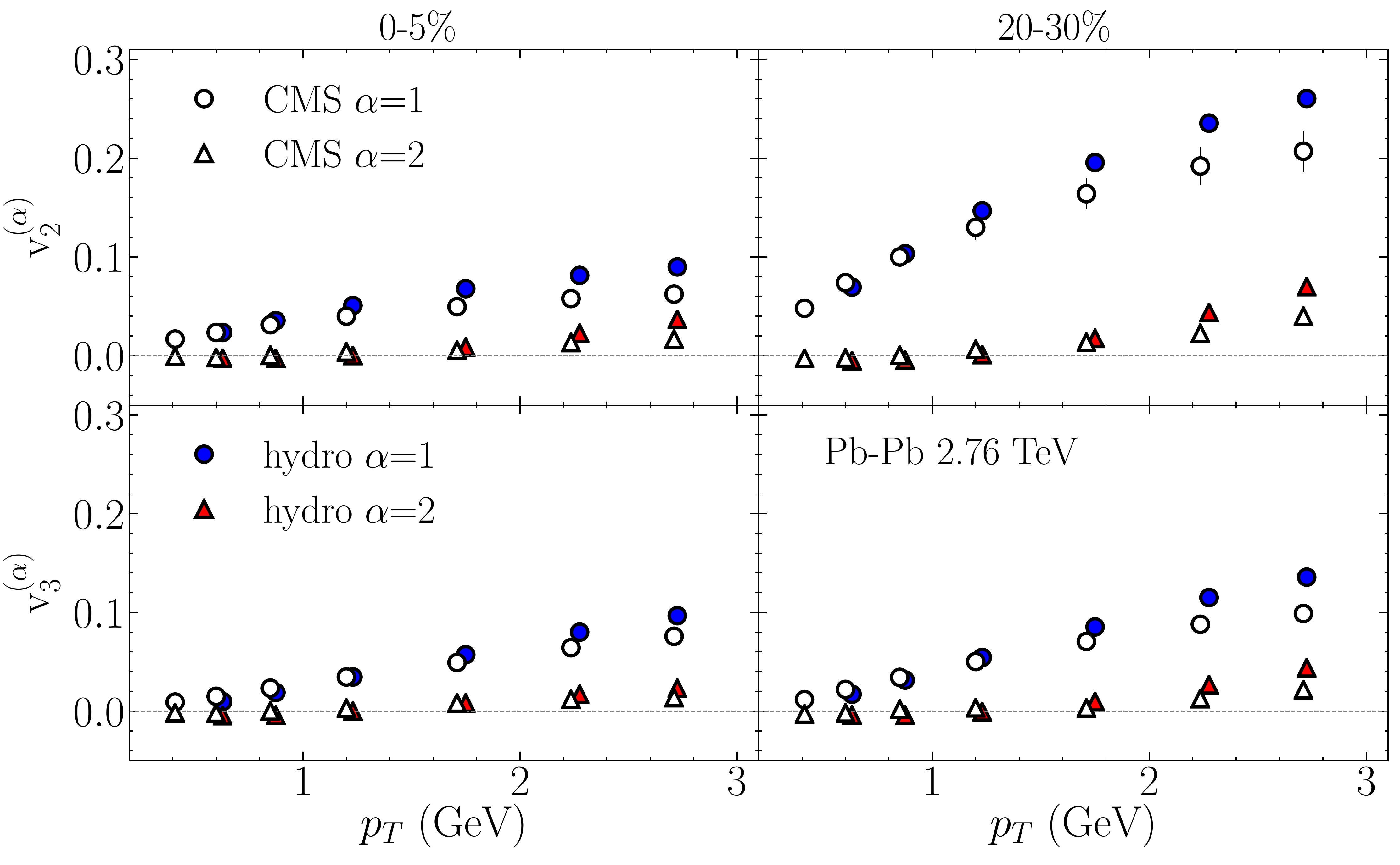}
\caption{
First two scaled principal components from the ideal fluid calculation in two
centrality windows corresponding to central (left) and midcentral (right)
collisions. Top: elliptic flow ($n=2$). Bottom: triangular flow ($n=3$).
Experimental data are from the CMS collaboration~\cite{Sirunyan:2017gyb}
(n=2,3). Symbols corresponding to experiment and theory have been slightly
shifted left and right for the sake of readability.
}
\label{fig:PCAn} 
\end{figure*}

For $n$=2--3, we show the first two scaled principal components and comparison
with CMS data in Fig. \ref{fig:PCAn}. Our cuts are $|\eta|< 2.5$ (equivalent to
CMS) but $p_T>0.5$ GeV, slightly higher than CMS $p_T>0.3$ GeV. We used similar
$p_T$ bins as experimentally.

The leading component is straightforward to interpret~\cite{Bhalerao:2014mua,
Mazeliauskas:2015vea,Mazeliauskas:2015efa}. If it dominates, Eq.~(\ref{eq:pca})
yields $\mathcal V_{n\Delta}(p_T^a,p_T^b) \sim \mathcal V_n^{(1)}(p_T^a)
\mathcal V_n^{(1)}(p_T^b)$ i.e. there is flow factorization. The event flow
defined by Eq.~(\ref{eq:pcaV}) reduces to $ \mathcal V_n(p_T) \sim
\xi_n^{(1)}(p_T) \mathcal V_n^{(1)}(p_T)$, i.e., the leading component
corresponds to usual anisotropic flow.
Concentrating on the region from 0 to 2 GeV, we see that
our hydro simulation slightly overestimates the leading components. Inclusion of
viscosity would damp them and improve agreement with data, as explicitly shown
for the $p_T$-integrated $n=3$ leading component in
Ref.~\cite{Mazeliauskas:2015vea}.

Higher-order principal components encode the information about the momentum
dependence of flow fluctuations.
They are in particular responsible for the breaking of factorization of two-particle correlations~\cite{Gardim:2012im}.
This effect is often quantified using the factorization breaking ratio $r_n$~\cite{Khachatryan:2015oea}, which is a function of two variables $p_T^a$ and $p_T^b$.
Higher-order principal components express the same information in a simpler way, because they are functions of a single variable $p_T^a$.
We only show the subleading component. 
In the range 0 to 2 GeV, our simulations capture the main features of the data.
The subleading component changes sign as a function of $p_T$, which is imposed by orthogonality with the leading mode. 
The fact that our $p_T$ cut is slightly higher than in data shifts this crossing point to the right. 
Note that inclusion of viscosity is not expected to change significantly the magnitude of the subleading mode, as was shown for $n=3$ in Ref.~\cite{Mazeliauskas:2015vea}.

\section{Results for multiplicities}\label{sec:mult}

\begin{figure*}
\includegraphics[width=0.8\textwidth]{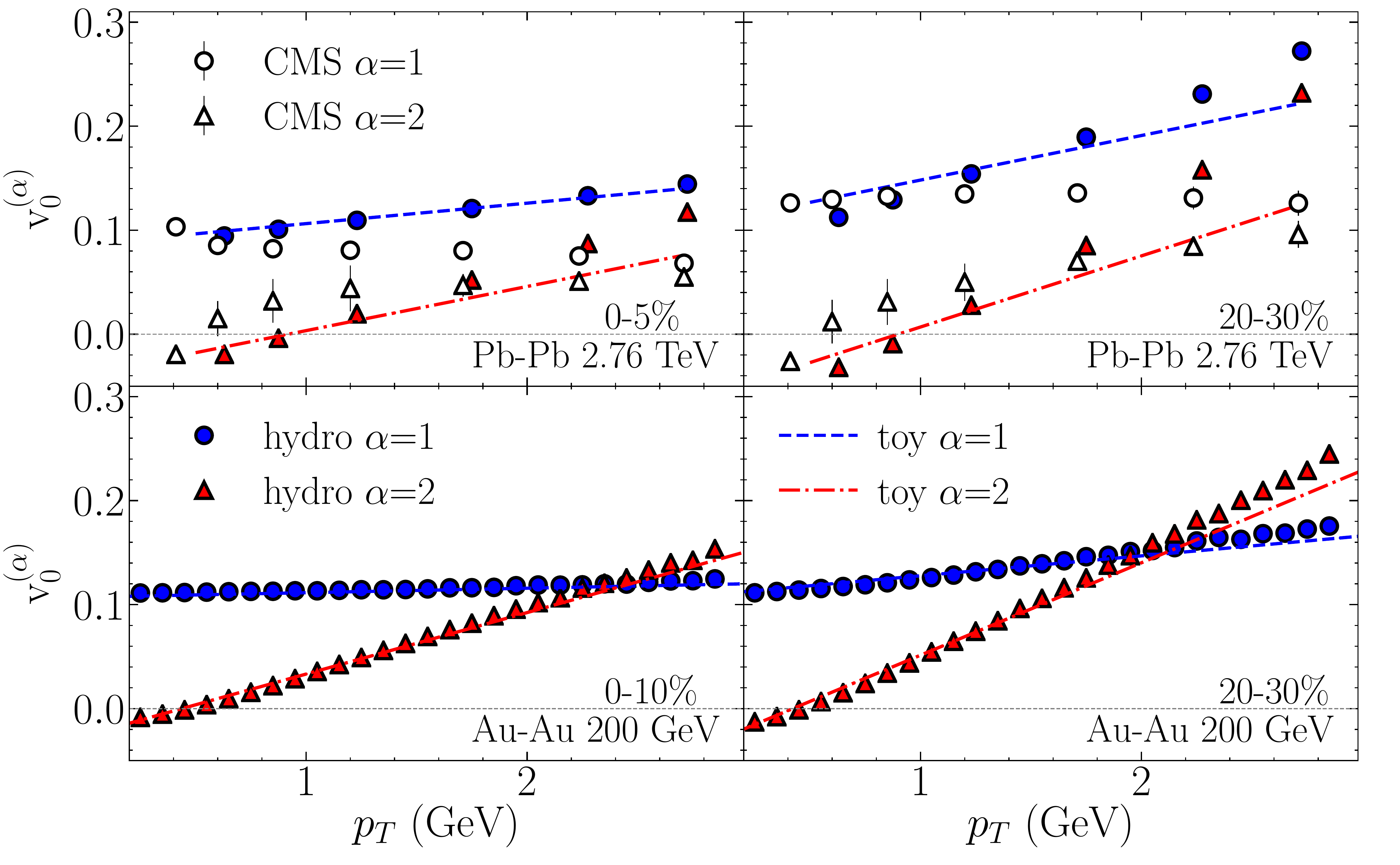} 
\caption{
First two scaled principal components for $n=0$ (multiplicity fluctuations). The
top panels display a comparison between our ideal fluid calculation, CMS
data~\cite{Sirunyan:2017gyb} and the approximate result from the toy model,
Eq.~(\ref{eq:pptoy}) (lines). The bottom panels display our predictions and the
toy model for Au+Au collisions at 200~GeV. As in Figs.~\ref{fig:PCAn}, the left
panels correspond to central collisions, and the right panels to mid-central
collisions.
}
\label{fig:PCA0} 
\end{figure*}

We now discuss multiplicity fluctuations, corresponding to $n=0$ principal
components. The comparison between our results and CMS data is displayed in the
top panels of Fig.~\ref{fig:PCA0}. There is rough overall agreement, but not as
good as in Fig.~\ref{fig:PCAn}. The leading component is rather independent of
$p_T$ in experiment, while it increases with $p_T$ in our hydrodynamic
calculation. The increase is less strong at RHIC energies (bottom panel of
Fig.~\ref{fig:PCA0}). The increase at LHC energies is not specific to our
implementation, as it has been seen by other groups~\cite{Mazeliauskas:2015efa,
Matt}. Such qualitative disagreement between hydrodynamics and experimental data
is rare, therefore, we investigate its origin in detail.\footnote{Note that the
transport model AMPT without hydrodynamics predicts a flat leading component, as
seen in data.}

In order to understand the principal components for $n=0$, we introduce a toy
model where the fluctuation of the multiplicity in a $p_T$ bin originates from
two sources:
1) fluctuations of the total multiplicity $N$.
2) fluctuations of the mean transverse momentum $\bar{p}_T$.
We assume that the $p_T$ spectrum is exponential:
\begin{equation}
    \label{spectra}
    \frac{1}{2\pi}\frac{dN}{dydp_T} =
    \mathcal V_0(p_T) =
    \frac{2 p_T N}{\pi{\bar{p}_T^2}} e^{-\frac{2 p_T}{\bar{p}_T}}
\end{equation}
where $N$ is the total multiplicity per unit rapidity and $\bar{p}_T$ is the
mean transverse momentum in one event. Next, we allow $N$ and $\bar{p}_T$ in a
given event to deviate from the event-averaged total multiplicity
$\langle N \rangle$, and the event-averaged mean transverse momentum
$\langle \bar p_T \rangle$ in a centrality bin, respectively:
\begin{align}
    N & = \langle N \rangle + \delta N,
    \\
    \bar p_T & = \langle \bar p_T \rangle + \delta \bar p_T.
\end{align}
Expanding Eq.~(\ref{spectra}) to first order in $\delta N$ and $\delta\bar p_T$,
one obtains:
\begin{equation}
    \label{fluct}
    \frac{\delta \mathcal V_0(p_T)}{\langle \mathcal V_0(p_T) \rangle} =
    \frac{\delta N}{\langle N \rangle} -
    2 \frac{\delta\bar p_T}{\langle \bar p_T \rangle} +
    2 \frac{p_T\delta\bar p_T}{\langle \bar p_T \rangle^2}.
\end{equation}
The covariance (\ref{defvndelta}) is then given by
\begin{equation}
    \label{defv0delta}
    \mathcal V_{0\Delta}(p_T^a,p_T^b) \equiv
    \langle \delta \mathcal V_0(p_T^a)\delta \mathcal V_0(p_T^b) \rangle,
\end{equation}
where angular brackets denote an average over events in a centrality bin.
Inserting Eq.~(\ref{fluct}) into Eq.~(\ref{defv0delta}), one obtains:
\begin{eqnarray}
    \label{v0delta}
    \frac{
        \mathcal V_{0\Delta}(p_T^a,p_T^b)
    }{
        \langle \mathcal V_0(p_T^a) \rangle \langle \mathcal V_0(p_T^b) \rangle
    }
    & = &
    \frac{\sigma_N^2}{\langle N \rangle^2} +
    4 \frac{\sigma_{p_T}^2}{\langle \bar{p}_T \rangle^2} -
    4 \frac{
        \left\langle \delta N \delta\bar p_T \right\rangle
    }{
        \langle N\rangle \langle \bar{p}_T\rangle
    }\cr
    & & +
    2 \left(
        \frac{
            \left\langle \delta N\delta\bar p_T \right\rangle
        }{
            \langle N\rangle \langle \bar{p}_T\rangle
        } -
        2 \frac{\sigma_{p_T}^2}{\langle \bar{p}_T\rangle^2}
    \right) \frac{p_T^a + p_T^b}{\langle \bar{p}_T \rangle}\cr
    & & +
    4 \frac{\sigma_{p_T}^2}{\langle \bar{p}_T \rangle^2}
        \frac{p_T^ap_T^b}{\langle \bar{p}_T \rangle^2},
\end{eqnarray}
where $\sigma_N^2 \equiv \langle \delta N^2 \rangle$ and $\sigma_{p_T}^2 \equiv
\langle \delta \bar p_T^2 \rangle$ denote the variance of the multiplicity and
mean $p_T$, respectively. Inspection of the dependence on $p_T^a$ and $p_T^b$
shows that the scaled principal components defined by Eqs.~(\ref{eq:pca}) and
(\ref{defscaled}) can only be of the form
\begin{equation}
    \label{2dim}
    v_0^{(\alpha)}(p_T) =
    a^{(\alpha)} + b^{(\alpha)} \frac{p_T}{\langle \bar p_T \rangle},
\end{equation}
i.e., they are linear in $p_T$. Since they span a two-dimensional space, this in
turn implies that there are at most two principal components (remember that
principal components are mutually orthogonal). The full analytic expressions of
these principal components are cumbersome. Therefore, we make further
simplifying assumptions, by identifying the leading terms in
Eq.~(\ref{v0delta}).

\begin{table}
\centering
\begin{tabular}{|c|c|c|c|c|c|}
\hline 
energy&centrality && $\frac{\sigma_N}{\langle N \rangle}$ &
$\frac{\sigma_{p_T}}{\langle \bar{p}_T \rangle}$ &
$\sqrt{\frac{
    \langle \delta N \delta \bar{p}_T \rangle
}{
    \langle N \rangle \langle \bar{p}_T \rangle
}}$ \\
\hline
2.76~TeV &0-5~\% &hydro& 0.12 & 0.026 & 0.041\\
&         & {\it CMS}      &{\it 0.09}& {\it 0.010}& {\it 0.\phantom{000}}\\
&20-30~\% &hydro& 0.16& 0.041 & 0.070\\
&         &   {\it CMS}    &{\it 0.13}& {\it 0.019}& {\it 0.020}\\
200~GeV&0-10~\% &hydro&0.11 & 0.017 & 0.017 \\
&20-30\%&hydro& 0.12 & 0.025 & 0.031 \\
\hline
\end{tabular}
\caption{
Values of the variances and covariance of $N$ and $\bar{p}_T$ at LHC and
RHIC in our hydrodynamical calculation using NeXSPheRIO.
The number in italics are approximate values extracted from CMS data through a rough fit of the principal components $v_0^{(1)}(p_T)$ and $v_0^{(2)}(p_T)$, shown in the top panels of Fig.~\ref{fig:PCA0}, using Eq.~(\ref{eq:pptoy}). 
}
\label{table:1}
\end{table}
Table~\ref{table:1} gives the values of the relative fluctuations of $N$ and
$\bar p_T$ in our hydrodynamic calculation, as well as their covariance. The
relative fluctuations of $N$ are larger by an order of magnitude, which is
explained by the large width of the centrality bin. In the limit where
$\sigma_{p_T}$ and $\langle\delta N\delta\bar{p}_T\rangle$ can be neglected,
only the first term remains in the right-hand side of Eq.~(\ref{v0delta}). The
covariance matrix trivially factorizes, i.e., there is only one principal
component. The scaled principal component, defined by Eq.~(\ref{defscaled}), is:
\begin{equation}
     v_0^{(1)}(p_T) \simeq \frac{\sigma_N}{\langle N \rangle}.
    \label{order0}
\end{equation}
It is independent of $p_T$. Thus, the fact that our hydrodynamic calculation
reproduces the magnitude of $v_0^{(1)}(p_T)$ at low $p_T$ (i.e., for the bulk of
produced particles) simply means that it has the correct multiplicity
fluctuations. These are largely dominated by the width of the centrality bin
used for the analysis, or, equivalently, by impact parameter fluctuations.

We now consider the more general case where
$\sigma_{p_T} / \langle \bar{p}_T \rangle$ and $\langle \delta N
\delta\bar{p}_T \rangle / \langle N \rangle \langle \bar{p}_T \rangle$ are not
zero, but can still be treated as small quantities. Then, to leading order in
these quantities, the scaled principal components are:
\begin{eqnarray}
    v_0^{(1)}(p_T) & \simeq &
    \frac{\sigma_N}{\langle N \rangle} +
    \left[\frac{
        - \left(\frac{\sigma_{p_T}}{\langle \bar{p}_T \rangle}\right)^2 +
          2 \frac{
              \langle \delta N \delta\bar p_T \rangle
          }{
              \langle N \rangle \langle \bar{p}_T \rangle
          }
      }{
          \left(\frac{\sigma_N}{\langle N \rangle}\right)
      }
   \right] \frac{p_T}{\langle \bar{p}_T \rangle}
    ,\cr
    v_0^{(2)}(p_T) & \simeq &
    - \frac{3}{2} \frac{\sigma_{p_T}}{\langle \bar{p}_T \rangle} \left(
        1 - \frac{4}{3}\frac{p_T}{\langle \bar{p}_T \rangle}
    \right).
    \label{eq:pptoy}
\end{eqnarray}
One can check that with these expressions, the decomposition (\ref{eq:pca}) is
satisfied. In terms of the scaled components, this equation can be written:
\begin{equation}
    \label{dec}
    \frac{
        \mathcal V_{0\Delta}(p_T^a,p_T^b)
    }{
        \langle \mathcal V_0(p_T^a) \rangle \langle \mathcal V_0(p_T^b) \rangle
    }
    = v_0^{(1)}(p_T^a) v_0^{(1)}(p_T^b) + v_0^{(2)}(p_T^a) v_0^{(2)}(p_T^b).
\end{equation}
Inserting Eq.~(\ref{eq:pptoy}) into Eq.~(\ref{dec}), and expanding to first
order in $\langle \delta N\delta\bar{p}_T \rangle$ and $\sigma_{p_T}^2$, one
recovers Eq.~(\ref{v0delta}) except for the second and third terms of the first
line, which are subleading corrections to the first term.

Equation~(\ref{eq:pptoy}) is a refinement of the zeroth-order result,
Eq.~(\ref{order0}). A subleading mode $v_0^{(2)}(p_T)$ appears, which is
directly proportional to $\sigma_{p_T} / \langle \bar{p}_T \rangle$. The
connection between the subleading mode and $p_T$ fluctuations was already made
in Ref.~\cite{Mazeliauskas:2015efa}. The 
change of sign of the subleading mode occurs at $p_T=(3/4)\langle \bar{p}_T \rangle$, which depends little on centrality.
Figure~\ref{fig:PCA0} displays a comparison between Eq.~(\ref{eq:pptoy}) and the
result from the full hydrodynamic calculation. Agreement is very good at RHIC
and a little worse at LHC (presumably due to the different lower $p_T$ cuts).
We therefore conclude that Eq.~(\ref{eq:pptoy}) captures the physics of
the first two $n=0$ modes. 

Using CMS data on principal components,
shown in Fig.~\ref{fig:PCA0}, one can estimate the quantities appearing in the right-hand side of Eq.~(\ref{eq:pptoy}). 
The corresponding numbers are reported in Table~\ref{table:1}, and should be considered rough figures. 
As explained above, the value of $\sigma_N/\langle N\rangle$ is given by the value of $v_0^{(1)}(p_T)$ at low $p_T$. 
The value of $\langle p_T\rangle$ is inferred from the value of $p_T$ for which $v_0^{(2)}(p_T)$ crosses the horizontal line.
This gives $\langle p_T\rangle\sim 0.75~GeV$, in reasonable agreement with the value $0.81~GeV$ obtained by direct integration of $p_T$-spectra in the same range (0.3 to 3 GeV)~\cite{Acharya:2018qsh}. 
The value of $\sigma_{p_T}/\langle\bar p_T\rangle$ is then estimated by fitting $v_0^{(2)}(p_T)$ at low $p_T$, and the resulting values agree with those from a dedicated analysis~\cite{Abelev:2014ckr}. 
The covariance $\langle \delta N\delta\bar p_T\rangle/\langle N\langle\langle\bar p_T\rangle$ is finally inferred from the $p_T$ dependence of $v_0^{(1)}(p_T)$. 
While the values of $\sigma_N/\langle N\rangle$ from the hydrodynamic calculation are in reasonable agreement with data, values of $\sigma_{p_T}/\langle\bar p_T\rangle$ are too large by a factor $\sim 2$, and the discrepancy is even worse for the covariance. 
We come back to this point below. 

The motivation for building the toy model was to understand under which
condition the leading mode is independent of $p_T$, or rises with $p_T$. The
first line of Eq.~(\ref{eq:pptoy}) shows that a rise with $p_T$ can be ascribed
to a positive correlation between the mean transverse momentum and the
multiplicity, represented by the quantity
$\langle \delta N \delta\bar{p}_T \rangle$. The fact that this rise is seen in
hydrodynamic calculations, not in data, implies that hydrodynamic calculations
overestimate $\langle \delta N \delta\bar{p}_T \rangle$, as illustrated by the numbers in Table~\ref{table:1} (the covariance extracted from CMS data for central collisions is compatible with 0). 
This can be related to
the fact that hydrodynamic models yield too large $\delta\bar p_T$ in general,
as pointed out by a study of transverse momentum
fluctuations~\cite{Bozek:2012fw}. Since transverse momentum fluctuations in
hydrodynamics originate from fluctuations in the transverse size of the
interaction region~\cite{Broniowski:2009fm},\footnote{At a given centrality, a smaller size implies a larger density and temperature, hence a larger mean transverse momentum.}, this in turn implies that existing
models of initial fluctuations tend to overestimate the size fluctuations.

The conclusion of this study is that a model which predicts the right
multiplicity and $p_T$ fluctuations should capture the first two principal
components for $n=0$. The reason why our hydrodynamical model predicts a rise of
the leading mode with $p_T$, which is not seen in data, can be related to the fact that that $\sigma_{p_T}$ is larger in our model than in data. 
Despite the fact that our hydrodynamic calculation at 2.76~TeV does not reproduce CMS data, we expect our predictions for 200~GeV collisions, shown in the bottom panels of Fig.~\ref{fig:PCA0}, should correctly predict the first two modes of multiplicity fluctuations at RHIC.
The reason is that the values of $\sigma_N/\langle N\rangle$
from Table~\ref{table:1} are comparable with experimental values from
PHENIX~\cite{Adare:2008ns}, and the values of $\sigma_{p_T} /
\langle \bar{p}_T \rangle$ are slightly too large compared to STAR
data~\cite{Adams:2005ka}, but in fair agreement.

\section{Conclusion}

We have compared results from a hydrodynamic simulation using the code
NeXSPheRIO with recent experimental data by CMS, on the Principal Component
Analysis. The trends for the leading and subleading components of elliptic and
triangular flow are in fair agreement with data. In contrast, for multiplicity
fluctuations, we have pointed out a qualitative disagreement: The leading
component increases with $p_T$ in hydrodynamics (here as well as in
\cite{Mazeliauskas:2015efa,Matt}) while it is constant in data at LHC energies.
We have constructed a toy model which gives result in good agreement with the
full hydrodynamic calculation. In this toy model, the subleading component is
proportional to the standard deviation of the mean $p_T$, $\sigma_{p_T}$. The
leading component is close to $\sigma_N/\langle N\rangle$ at low $p_T$, but
increases with $p_T$ if the fluctuations of $p_T$ are large and correlated with
the fluctuations of the multiplicity.

We have thus related $n=0$ results from the principal component analysis to
multiplicity and transverse momentum fluctuations. Fluctuations in $N$ and
$\bar{p}_T$ have been attracting attention for a long time because they may
probe the QCD phase transition (see e.g. \cite{Stephanov:2005iu}), as well as
initial inhomogeneities (see for example \cite{Gavin:2011gr,Broniowski:2009fm}).
The principal components are sensitive not only to the width of multiplicity and
transverse momentum fluctuations, but also to their mutual covariance. They open
a new window on initial fluctuations, which can be used to rule out initial
condition models.

\section*{Acknowledgments}

F.G.~acknowledges support from Funda\c{c}\~ao de Amparo \`a Pesquisa do Estado
de S\~ao Paulo (FAPESP grant 2018/18075-0) and Conselho Nacional de
Desenvolvimento Cient\'{\i}fico e Tecnol\'ogico (CNPq grant 310141/2016-8).
F.G.G. was supported by Conselho Nacional de Desenvolvimento Cient\'{\i}fico e
Tecnol\'ogico (CNPq grant 205369/2018-9 and 312932/2018-9) and FAPEMIG (grant
APQ-02107-16). P.I.~thanks support from Coordena\c{c}\~ao de Aperfei\c{c}oamento
de Pessoal de N\'{\i}vel Superior (CAPES) and Conselho Nacional de
Desenvolvimento Cient\'{\i}fico e Tecnol\'ogico (CNPq). M.L.~acknowledges
support from FAPESP projects 2016/24029-6 and 2017/05685-2. F.G., F.G.G. and
M.L. acknowledge support from project INCT-FNA Proc.~No.~464898/2014-5 and F.G.,
M.L. and J.-Y.O.  from USP-COFECUB (grant Uc Ph 160-16, 2015/13).


\begin{thebibliography}{99}

\bibitem{Gardim:2011xv} 
F.~G.~Gardim, F.~Grassi, M.~Luzum and J.~Y.~Ollitrault,
Phys.\ Rev.\ C {\bf 85}, 024908 (2012).



\bibitem{Gardim:2014tya} 
F.~G.~Gardim, J.~Noronha-Hostler, M.~Luzum and F.~Grassi,
Phys.\ Rev.\ C {\bf 91}, no. 3, 034902 (2015).



\bibitem{Niemi:2012aj} 
H.~Niemi, G.~S.~Denicol, H.~Holopainen and P.~Huovinen,
Phys.\ Rev.\ C {\bf 87}, no. 5, 054901 (2013).



\bibitem{Niemi:2015qia} 
H.~Niemi, K.~J.~Eskola and R.~Paatelainen,
Phys.\ Rev.\ C {\bf 93}, no. 2, 024907 (2016).



\bibitem{Fu:2015wba} 
J.~Fu,
Phys.\ Rev.\ C {\bf 92}, no. 2, 024904 (2015).



\bibitem{Noronha-Hostler:2015dbi} 
J.~Noronha-Hostler, L.~Yan, F.~G.~Gardim and J.~Y.~Ollitrault,
Phys.\ Rev.\ C {\bf 93}, no. 1, 014909 (2016).



\bibitem{Gardim:2012im} 
F.~G.~Gardim, F.~Grassi, M.~Luzum and J.~Y.~Ollitrault,
Phys.\ Rev.\ C {\bf 87}, no. 3, 031901 (2013).



\bibitem{Kozlov:2014fqa} 
I.~Kozlov, M.~Luzum, G.~Denicol, S.~Jeon and C.~Gale,
arXiv:1405.3976 [nucl-th].



\bibitem{Heinz:2013bua} 
U.~Heinz, Z.~Qiu and C.~Shen,
Phys.\ Rev.\ C {\bf 87}, no. 3, 034913 (2013).



\bibitem{Khachatryan:2015oea} 
V.~Khachatryan {\it et al.} [CMS Collaboration],
Phys.\ Rev.\ C {\bf 92}, no. 3, 034911 (2015).



\bibitem{Gardim:2017ruc} 
F.~G.~Gardim, F.~Grassi, P.~Ishida, M.~Luzum, P.~S.~Magalhães and
J.~Noronha-Hostler,
Phys.\ Rev.\ C {\bf 97}, no. 6, 064919 (2018).



\bibitem{Zhao:2017yhj} 
W.~Zhao, H.~j.~Xu and H.~Song,
Eur.\ Phys.\ J.\ C {\bf 77}, no. 9, 645 (2017).



\bibitem{McDonald:2016vlt} 
S.~McDonald, C.~Shen, F.~Fillion-Gourdeau, S.~Jeon and C.~Gale,
Phys.\ Rev.\ C {\bf 95}, no. 6, 064913 (2017).



\bibitem{Bhalerao:2014mua} 
R.~S.~Bhalerao, J.~Y.~Ollitrault, S.~Pal and D.~Teaney,
Phys.\ Rev.\ Lett.\ {\bf 114}, no. 15, 152301 (2015).



\bibitem{Mazeliauskas:2015vea} 
A.~Mazeliauskas and D.~Teaney,
Phys.\ Rev.\ C {\bf 91}, no. 4, 044902 (2015).



\bibitem{Mazeliauskas:2015efa} 
A.~Mazeliauskas and D.~Teaney,
Phys.\ Rev.\ C {\bf 93}, no. 2, 024913 (2016).



\bibitem{Sirunyan:2017gyb} 
A.~M.~Sirunyan {\it et al.} [CMS Collaboration],
Phys.\ Rev.\ C {\bf 96}, no. 6, 064902 (2017).



\bibitem{pcaref1} J. O. Ramsay and B. W. Silverman, 
``Functional Data Analysis'', Springer, New York-USA (2005).


\bibitem{Bozek:2017thv} 
P.~Bozek,
Phys.\ Rev.\ C {\bf 97}, no. 3, 034905 (2018).

\bibitem{Liu:2019jxg}
Z.~Liu, W.~Zhao and H.~Song, 
``Principal Component Analysis of collective flow in Relativistic Heavy-Ion
Collisions'', arXiv:1903.09833



\bibitem{Drescher:2000ha} 
H.~J.~Drescher, M.~Hladik, S.~Ostapchenko, T.~Pierog and K.~Werner,
Phys.\ Rept.\ {\bf 350}, 93 (2001).



\bibitem{Qian:2007ff} 
W.~L.~Qian, R.~Andrade, F.~Grassi, O.~Socolowski, Jr., T.~Kodama and Y.~Hama,
Int.\ J.\ Mod.\ Phys.\ E {\bf 16}, 1877 (2007).



\bibitem{Andrade:2006yh} 
R.~Andrade, F.~Grassi, Y.~Hama, T.~Kodama and O.~Socolowski, Jr.,
Phys.\ Rev.\ Lett.\ {\bf 97}, 202302 (2006).



\bibitem{Andrade:2008xh} 
R.~P.~G.~Andrade, F.~Grassi, Y.~Hama, T.~Kodama and W.~L.~Qian,
Phys.\ Rev.\ Lett.\ {\bf 101}, 112301 (2008).



\bibitem{Andrade:2008fa} 
R.~P.~G.~Andrade, A.~L.~V.~R.~dos Reis, F.~Grassi, Y.~Hama, W.~L.~Qian,
T.~Kodama and J.-Y.~Ollitrault,
Acta Phys.\ Polon.\ B {\bf 40}, 993 (2009).



\bibitem{Gardim:2011qn} 
F.~G.~Gardim, F.~Grassi, Y.~Hama, M.~Luzum and J.~Y.~Ollitrault,
Phys.\ Rev.\ C {\bf 83}, 064901 (2011).



\bibitem{Gardim:2012yp} 
F.~G.~Gardim, F.~Grassi, M.~Luzum and J.~Y.~Ollitrault,
Phys.\ Rev.\ Lett.\ {\bf 109}, 202302 (2012).



\bibitem{Takahashi:2009na}
J.~Takahashi, B.~M.~Tavares, W.~L.~Qian, R.~Andrade, F.~Grassi, Y.~Hama,
T.~Kodama and N.~Xu,
Phys.\ Rev.\ Lett.\ {\bf 103}, 242301 (2009).



\bibitem{Qian:2012qn}
W.~L.~Qian, R.~Andrade, F.~Gardim, F.~Grassi and Y.~Hama,
Phys.\ Rev.\ C {\bf 87}, no. 1, 014904 (2013).



\bibitem{RHICdist}
L.~Barbosa, F.~Gardim, F.~Grassi, M.~Luzum and M.~V.~Machado, in preparation.

\bibitem{Matt}
M. ~Luzum, private communcation about MUSIC code with Trento initial conditions.

%
%
%

\bibitem{Acharya:2018qsh} 
  S.~Acharya {\it et al.} [ALICE Collaboration],
  JHEP {\bf 1811}, 013 (2018)
  doi:10.1007/JHEP11(2018)013
  [arXiv:1802.09145 [nucl-ex]].

\bibitem{Abelev:2014ckr} 
  B.~B.~Abelev {\it et al.} [ALICE Collaboration],
  Eur.\ Phys.\ J.\ C {\bf 74}, no. 10, 3077 (2014)
  doi:10.1140/epjc/s10052-014-3077-y
  [arXiv:1407.5530 [nucl-ex]].


\bibitem{Bozek:2012fw} 
P.~Bozek and W.~Broniowski,
Phys.\ Rev.\ C {\bf 85}, 044910 (2012).



\bibitem{Broniowski:2009fm} 
W.~Broniowski, M.~Chojnacki and L.~Obara,
Phys.\ Rev.\ C {\bf 80}, 051902 (2009).



\bibitem{Stephanov:2005iu} 
M.~Stephanov,
J.\ Phys.\ Conf.\ Ser.\ {\bf 27}, 144 (2005).



\bibitem{Gavin:2011gr} 
S.~Gavin and G.~Moschelli,
Phys.\ Rev.\ C {\bf 85}, 014905 (2012).



\bibitem{Adare:2008ns} 
A.~Adare {\it et al.} [PHENIX Collaboration],
Phys.\ Rev.\ C {\bf 78}, 044902 (2008).



\bibitem{Adams:2005ka} 
J.~Adams {\it et al.} [STAR Collaboration],
Phys.\ Rev.\ C {\bf 72}, 044902 (2005).



 
\end{thebibliography}
\end{document}